\documentclass[preprint,12pt]{elsarticle}

\usepackage{amssymb}

\usepackage{hyperref}

\journal{Applied Surface Science}

\begin{document}

\begin{frontmatter}



\title{Electrochemical Polishing of Chemical Vapor Deposited Niobium Thin Films}


\author[inst1]{Zeming Sun\corref{firstcorr}}
\cortext[firstcorr]{zs253@cornell.edu}
\author[inst1]{Mingqi Ge\fnref{firstnewadd}}
\fntext[firstnewadd]{Now at Jefferson Lab}
\author[inst1]{James T. Maniscalco\fnref{secnewadd}}
\fntext[secnewadd]{Now at SLAC}
\author[inst2]{Victor Arrieta}
\author[inst2]{Shawn R. McNeal}
\author[inst1]{Matthias U. Liepe\corref{secondcorr}}
\cortext[secondcorr]{mul2@cornell.edu}

\affiliation[inst1]{organization={Cornell Laboratory for Accelerator-Based Sciences and Education},
            city={Ithaca},
            postcode={14853}, 
            state={NY},
            country={USA}}

\affiliation[inst2]{organization={Ultramet},
            city={Pacoima},
            postcode={12173}, 
            state={CA},
            country={USA}}

\begin{abstract}
Combining chemical vapor deposition (CVD) with electrochemical polish (EP) operations is a promising route to producing performance-capable superconducting films for use in the fabrication of cost-effective components for superconducting radiofrequency (SRF) particle accelerators and superconducting quantum computers. The post-deposition EP process enables a critically necessary reduction in surface roughness of niobium thin films to promote optimal superconducting surface conditions. In this work, surface morphology, roughness, and crystal orientation of the CVD-grown and EP-polished niobium films were investigated. The grain growth and polishing mechanisms were analyzed. The CVD films were found to comprise steps, kinks, and pyramidal features, resulting in undesirable large peak-to-valley distances. The electrochemical polish was demonstrated to significantly diminish the height of pyramids and effectively minimize the overall surface roughness. In contrast to buffered chemical polishing (BCP), EP results showed a probable dependence on crystal orientation, suggesting this process was influenced by locally enhanced current density and thickness variations of oxide dielectrics. These understandings identify the EP principles tied to CVD-grown Nb films that allow further refinement of surface profiles for film-based SRF applications. 

\end{abstract}

\begin{graphicalabstract}
\includegraphics[width= 13 cm]{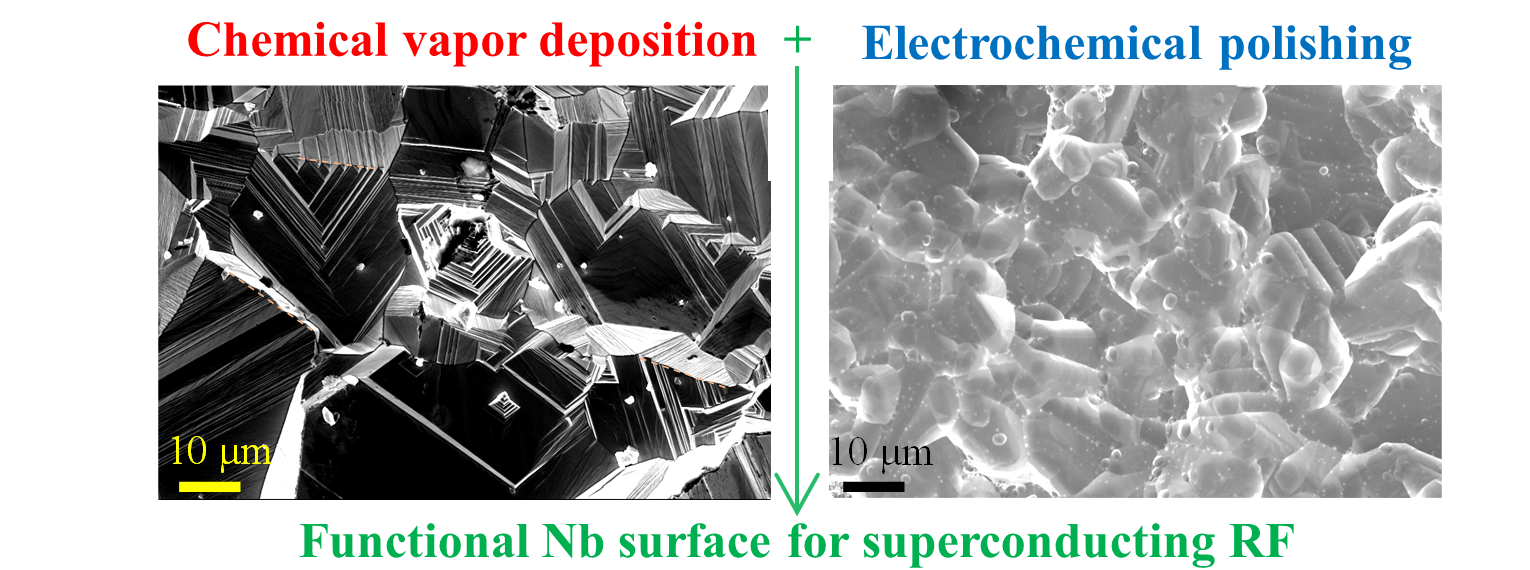}
\end{graphicalabstract}

\begin{highlights}
\item Electrochemical polishing (EP) is demonstrated to effectively minimize the surface roughness for chemical vapor deposited (CVD) niobium thin films.
\item CVD niobium films contain steps, kinks, and pyramidal features, resulting in large surface roughness. EP polishing of these films involves both macroscale and microscale smoothing.
\item A probable dependence on crystal orientation during EP is observed, indicating strong influences from locally enhanced current density and thickness variations of oxide dielectrics.
\item Obtaining the required surface conditions by a combined EP-CVD technology marks a feasible application of niobium thin films in superconducting RF.    
\end{highlights}


\end{frontmatter}


\section{Introduction}

Niobium (Nb) is an important superconducting material that finds use in superconducting radio-frequency (SRF) cavities, the chamber containing the electromagnetic field in modern particle accelerators \cite{SunRef1}, and in components needed in the emerging technological field of quantum computers \cite{SunRef2}. SRF cavities are critical components in a wide range of applications, including synchrotron and free-electron-laser light sources (\textit{e.g.}, Linac Coherent Light Source (LCLS)) \cite{SunRef3,SunRef4}, high energy physics such as in the search for dark matter \cite{SunRef5}, high-precision ($<$ 5 nm) photolithography for semiconductor device fabrication \cite{SunRef6}, and in biopharmaceutical and medical applications \cite{SunRef7}.

Since the transition of accelerators from low-gradient normal-conducting RF to high-gradient superconducting RF, bulk Nb remains as the dominant cavity technology used to obtain high accelerating gradients. Bulk Nb cavities are comprised of high-purity Nb with a residual resistivity ratio (RRR) exceeding 300 and require high-cost triple arc-melted RRR-500+ start materials for fabrication. One promising direction for realizing cost-effective cavities for SRF applications is the use of thin-film Nb coatings applied to low-cost, high-thermal-conducting copper (Cu) cavity substrates. The thin-film technology is viable since the active region for an SRF cavity is dictated by the field penetration depth, typically, tens to hundreds of nanometers at the inner surface, $\textit{e.g.}$, $\sim$ 40 nm for Nb. Additionally, due to the improved thermal conductance, the Nb-coated Cu cavity promises enhanced thermal stability during operation. The structural Cu cavity wall enables the outward diffusion and removal of waste heat, while the Nb film functions as the critical component interacting with the RF field. Controlling cavity surface roughness and mitigating surface defects are important for achieving high-quality factors as localized heat generated by these features can result in the cascading loss of the superconducting state on the cavity surface, an effect known as “quench” \cite{SunRef8}. 

\begin{figure}[!t]
\centering
\includegraphics[width= 13 cm]{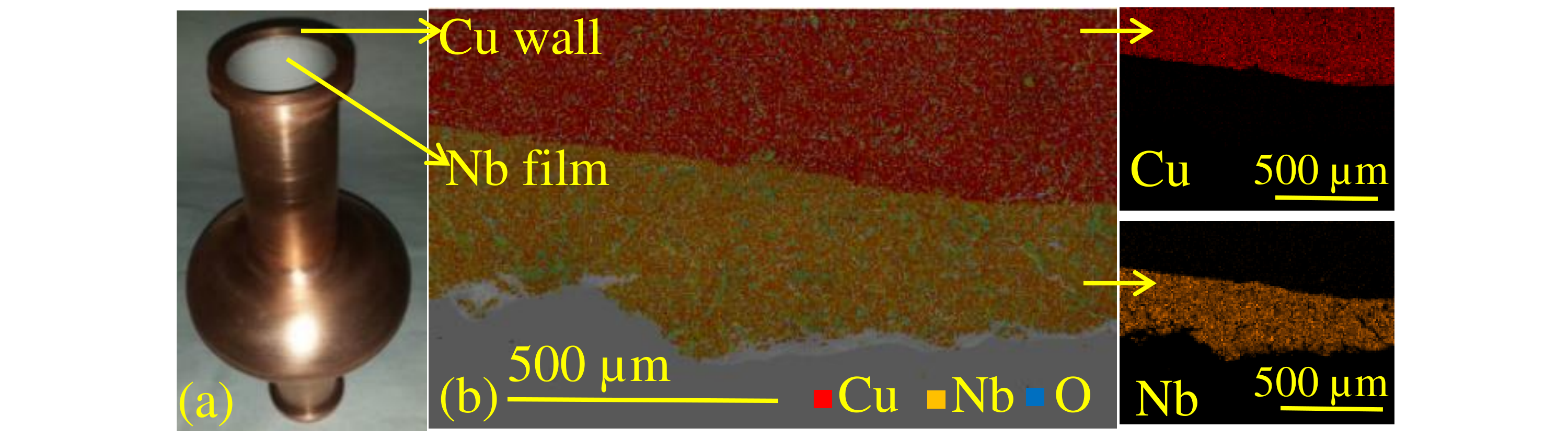}
\caption{(a) Picture of a Cu SRF cavity coated with CVD Nb thin films at the inner surface. (b) Cross-sectional EDS mapping of CVD Nb films on Cu. Samples were cut from the cavity. Inserts show locations of Cu substrate and Nb films.}
\label{SunFig1}
\end{figure}

Chemical vapor deposition (CVD) of Nb films, in addition to sputtering \cite{SunRef9,SunRef10,SunRef11} and epitaxy \cite{SunRef12}, were studied on silicon-carbide and graphite substrates using NbCl$_5$ and NbBr$_5$ precursors \cite{SunRef13,SunRef14,SunRef15}. This vapor-based technique is suitable for coating the inner surface of cavities with intricate shapes. Ultramet developed advanced CVD processing to deposit high-RRR ($>$ 280) and used rapid CVD process capabilities to produce freestanding testable bulk Nb 3.9 GHz cavities \cite{SunRef17}. Ultramet, working with Cornell’s SRF Group, adapted the advanced CVD process technology to vapor deposit thick-, and thin-film Nb on 5-inch diameter plates and then scaled the process to form Nb films on the interior surface of 1.3 GHz elliptical Cu cavities of the full-scale single-cell ILC design (Fig.~\ref{SunFig1}a) \cite{SunRef17,SunRef16}. Thin-film CVD Nb coatings produced by Ultramet in this work demonstrated a high-quality factor above 10$^{10}$ at 2 K and a low residual resistance of $\sim 5$ n$\Omega$ \cite{SunRef16}. Fig.~\ref{SunFig1}b shows the results of the elemental mapping via an energy-dispersive X-ray spectroscope (EDS), over the cross-section of a sample cut from the Nb/Cu cavity that had been electrochemically polished. The excellent Nb-Cu interface in the image confirms the $\sim$ 400 µm Nb film is strongly bonded to the Cu substrate, and no Cu inclusions are observed in the film. However, a large thickness variation of $\sim$ 150 µm remains even after the electrochemical polishing operation. The surface roughness can locally enhance the magnetic field and negatively impact the RF performance, due for example, to the degradation of quality factors (Q$_0$) at high accelerating gradients \cite{SunRef18}. Also, this type of field enhancement can cause a quench and limit the maximum field capability due to the permanent loss of superconductivity.

\begin{figure}[!t]
\centering
\includegraphics[width=\linewidth]{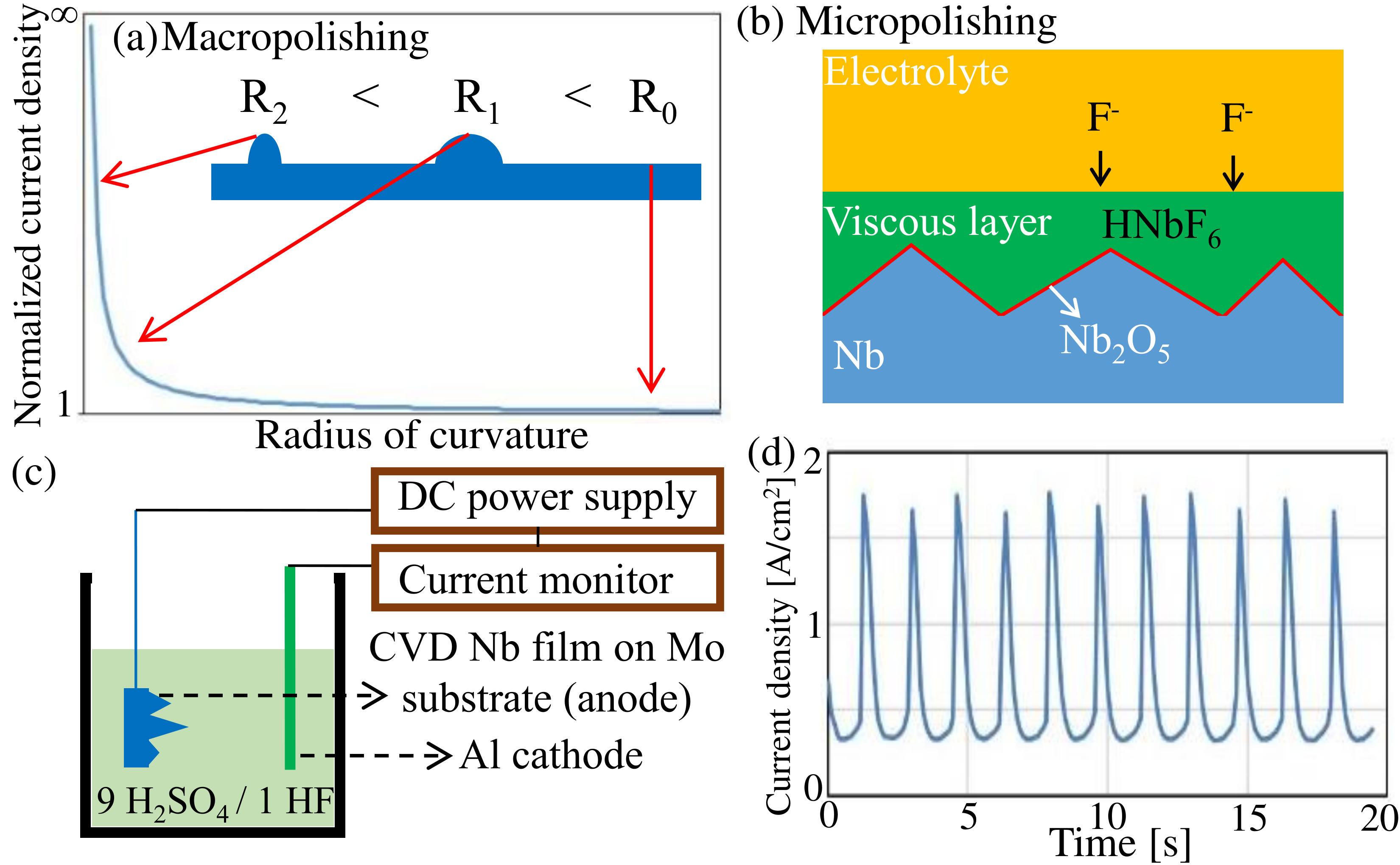}
\caption{(a,b) Mechanisms of electrochemical polishing on a niobium surface using H$_2$SO$_4$/HF electrolytes: (a) macropolishing and (b) micropolishing. (c) Schematic of the electrochemical polishing system and (d) polishing current oscillation. }
\label{SunFig2}
\end{figure}

As such, engineering a smooth RF surface is required. Previous investigations on bulk Nb involved mechanical polish \cite{SunRef19}, the use of chemicals such as buffered chemical polish (BCP) \cite{SunRef20}, and electrochemical polish (EP) \cite{SunRef21}. Among these methods, the EP process that employs 9-part concentrated H$_2$SO$_4$ to 1-part 48\% HF under a DC current is typically performed as a critical surface finish yielding an encouraging result of ~300 nm roughness on bulk Nb \cite{SunRef22}. A review of the literature suggests that an investigation into EP processing to condition Nb thin-film surfaces for SRF applications has not yet been done. 

Electrochemical polishing includes two categories regarding surface feature size, macropolishing and micropolishing. Landolt et al. \cite{SunRef23,SunRef24} and Hryniewicz et al. \cite{SunRef25} have reviewed the fundamental aspects of each. As shown in Fig.~\ref{SunFig2}a, the local current density is significantly enhanced at positions with a smaller radius of curvature as described via \cite{SunRef26}

\begin{equation}
    \sigma = \frac{\frac{2 \varepsilon \Delta V}{R}}{exp(\frac{-2\Delta n}{R})-1}_{\Delta~n \rightarrow 0}
\end{equation}
where $\sigma$ is the surface charge density, R is the radius of curvature, $\Delta$n is a limited distance normal to the surface, $\Delta$V is the potential difference between two endpoints of the distance $\Delta$n, and $\varepsilon$ is electric permittivity. Thus, for a surface with high roughness, the leveling of the peak and recessed regions via macropolishing is primarily determined by their difference in their current density. In contrast, a submicrometer-roughness surface has large radius-of-curvature features (closer to R$_0$ in Fig.~\ref{SunFig2}a), leading to a more uniform electrical field between peak and recessed regions, and making the micropolishing dominant by way of controlling the mass transport of species such as reactants (water, F$^{-}$, SO$_{4}^{2-}$) and products (HNbF$_6$ and other complexes). Numerous studies have been carried out to investigate the transport mechanism in play during polishing operations performed on bulk Nb surfaces \cite{SunRef21,SunRef27,SunRef28}. Tian et al. \cite{SunRef21,SunRef27} identified the limiting of the transport of F- ions as one mechanism and validated the theoretical interface model, as illustrated in Fig.~\ref{SunFig2}b, showing a compact Nb$_2$O$_5$ film and an HNbF$_6$ (and other complexes) diffusion layer. A viscous layer and/or dielectric film is formed between the bulk solid and liquid regions so that the reaction is facilitated at the peak region where random diffusion of species (F$^{-}$) is feasible as compared to the recessed region. 

Limitations in applying EP to thin Nb films arise due to the distinctive surface profile and structural properties induced by CVD, which are detailed in this work. For example, a variety of feature sizes appear on the film surface ranging from $\sim$ 100 $\mu$m, large pyramidal features to several nm-size kinks and steps, and present the challenge of smoothing the surface at the limit of allowed polish thickness. Moreover, crystal defects such as dislocations, impurities, and vacancies together with intrinsic stress in the film are more common than bulk Nb. Owing to the defective sites, there is concern over the formation of compact dielectric films as well as a desirable distribution of electric fields. Cu EP studies have reported failure of dielectric formation on a film sample and hence, a negative polish result, as compared to a bulk sample \cite{SunRef29}. These challenges motivate us to investigate EP on Nb thin films.

Here we analyze new phenomena tied to the EP treatment of CVD-grown Nb films and to further advance the EP-CVD combined technology, paving the way for film-based Nb RF cavities and other superconducting applications. We focus on comparing the characteristics between as-deposited and electrochemically polished films. Specifically, we investigate surface morphology, roughness, and grain orientation. Also, we discuss the CVD growth mode since these unique surface features observed are critical for determining the mechanism of a subsequent EP process. Moreover, the EP results to date indicate a probable dependence on crystal orientation; and analysis is provided in comparison with the chemically-controlled BCP treatment. 
\begin{figure}[!t]
\centering
\includegraphics[width=\linewidth]{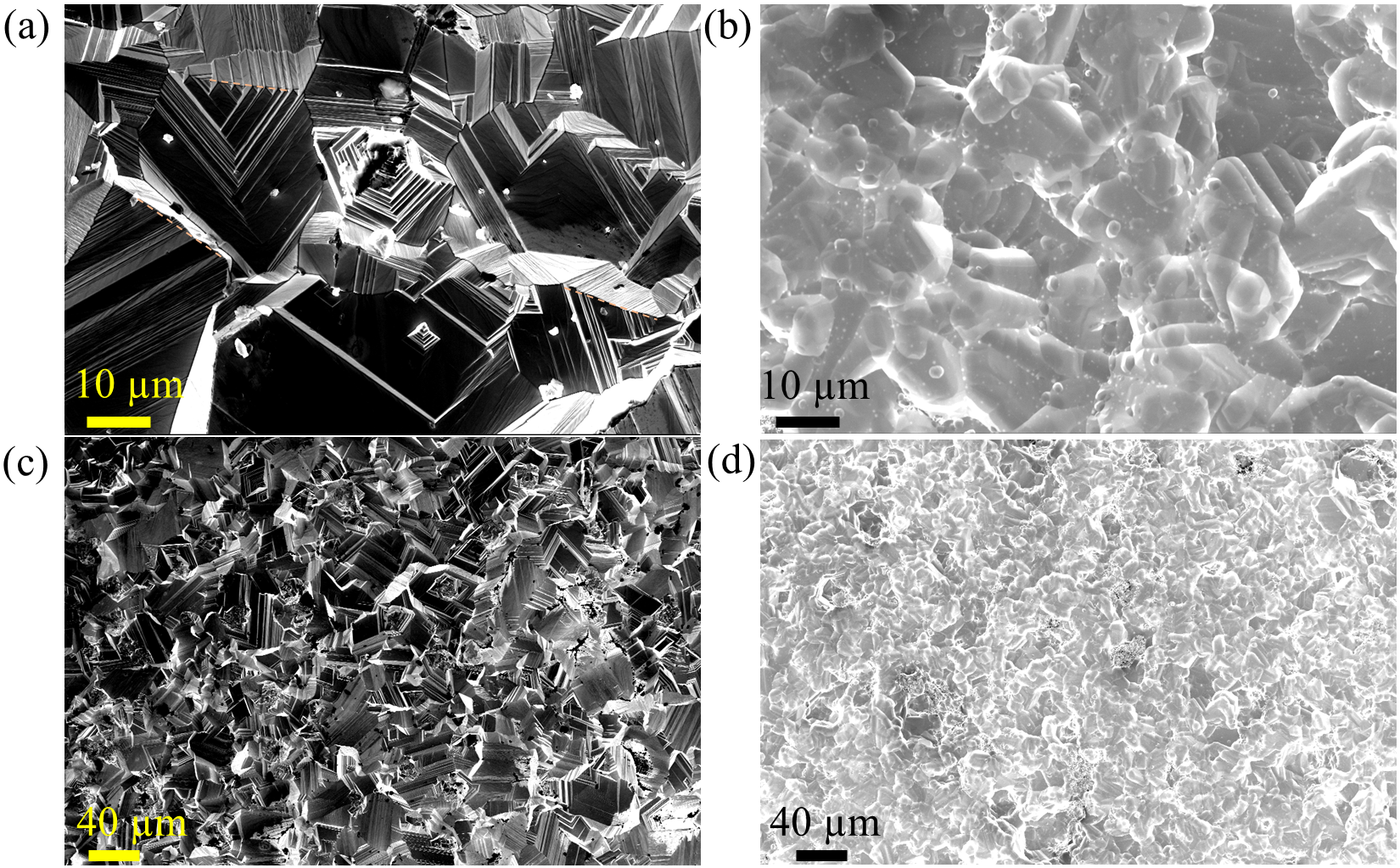}
\caption{Comparison of surface SEM images for CVD Nb films on the Mo substrate (a,c) before and (b,d) after EP under different fields of width: (a,b) 100 µm, (c,d) 500 µm.}
\label{SunFig3}
\end{figure}
\section{Experimental section}

Thin films ($>$ 100 $\mu$m) of Nb on the molybdenum (Mo) substrates were prepared by a low-temperature CVD process. The CVD Nb thin films were provided by Ultramet and the recipes are not disclosed. The as-deposited films were electrochemically polished by nominally 10 $\mu$m in thickness using a 2-electrode system (Fig.~\ref{SunFig2}c) consisting of the CVD Nb/Mo as an anode, Al as a cathode, and the electrolyte of 98\% H$_{2}$SO$_{4}$ and 48\% HF at a 9:1 volume ratio. The 2-electrode system is commonly used in the cavity polish at Cornell, FNAL, KEK, and other accelerator laboratories \cite{SunRef16,SunRef22,SunRef30}. The current oscillation regime (Fig.~\ref{SunFig2}d) was monitored to facilitate the generation and subsequent removal of compact Nb$_2$O$_5$ dielectrics. For reference to EP, samples were polished in a standard BCP (buffered chemical polishing) solution with 48\% hydrofluoric, 70\% nitric, and 85\% phosphoric acids at a volume ratio of 1:1:1. 

\begin{figure}[!t]
\centering
\includegraphics[width=\linewidth]{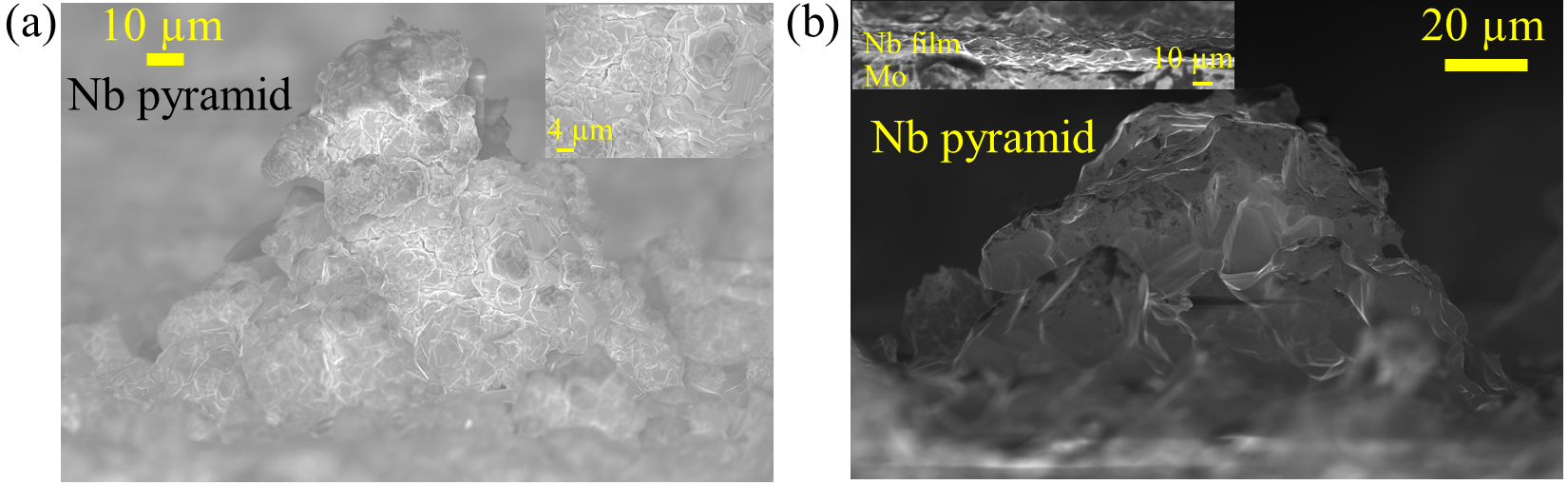}
\caption{Comparison of cross-sectional SEM images for the largest pyramidal features observed (a) before and (b) after EP. Inserts show closer inspections of (a) the CVD pyramid and (b) the relatively smooth regions after EP.}
\label{SunFig4}
\end{figure}

To evaluate the surface morphology change, surface and cross-sectional imaging were performed using a Zeiss Gemini scanning electron microscope (SEM) equipped with an in-lens detector under low voltage regimes (1 -- 5 kV). Electron dispersive x-ray spectroscopy (EDS) was used to determine the chemical information. The surface roughness of films was measured via an atomic force microscope (AFM, Asylum MFP-3D) but the high ($>$ 100 $\mu$m) pyramids affected the measurement, so the AFM results only compared the relatively smooth regions. To obtain effective comparison, films were vertically placed under the SEM, and the cross-sections of the highest pyramids were imaged and compared. Moreover, high-resolution X-ray diffraction (XRD, Rigaku SmartLab) patterns were collected for analyzing grain orientations. A Cu K$\alpha$ radiation with a wavelength of 0.154 nm was used. 

\section{Results and discussion}
\subsection{Surface morphology}

Fig.~\ref{SunFig3} shows the surface morphology of as-deposited and EP'ed films. As-deposited films (Fig.~\ref{SunFig3}a), although uniformly covering the substrate surface, exhibit features of facets and steps. Also notably, pyramid-like structures are widely observed on the surface as inspected under large fields of width (Fig.~\ref{SunFig3}c). The cross-section of the largest pyramid observed is presented in Fig.~\ref{SunFig4}a. To summarize, there are two sources of surface roughness: (1) pyramids as high as 100 $\mu$m; (2) step-kink structures appearing both in the relatively flat regions and on the pyramids. Note that small but sharp features, $\textit{e.g.}$, steps, would negatively affect the RF performance due to strong local field enhancement. Hence, polishing the film surface is necessary to improve the surface condition. 

\begin{figure}[!t]
\centering
\includegraphics[width= 7.62 cm]{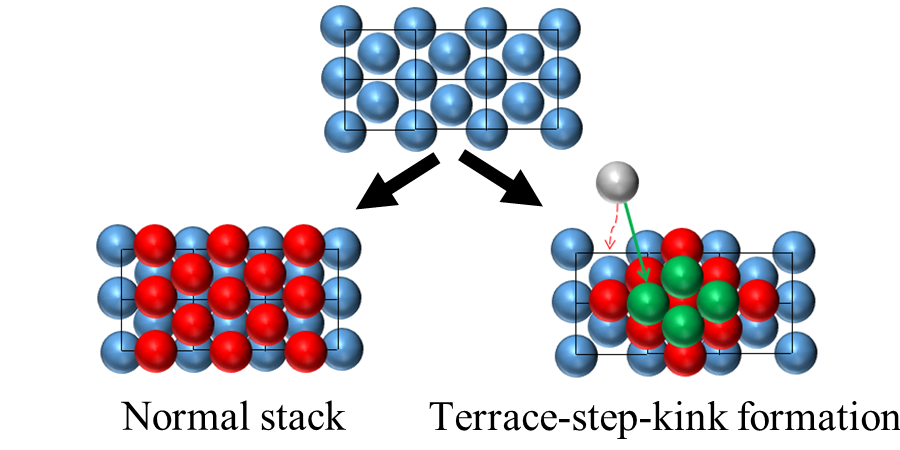}
\caption{Atomic models showing the terrace-step-kink formation on the Nb (110) plane. Blue, red, and green atoms indicate the 1st, 2nd, and 3rd atomic layers, respectively.}
\label{SunFig5}
\end{figure}

Regarding the step-kink and pyramid formation, we analyze the film growth mechanism. Based on a typical terrace-step-kink model \cite{SunRef31}, the nucleation events occur on multiple sites and a subsequent island growth mode forms the pyramid structure. As shown in Fig.~\ref{SunFig5}, the Nb atoms, as a result of the chemical reactions of precursors, are adsorbed on a terrace (the flat surface) and then diffuse to a kink site (the site at the terrace edge) where the surface energy is typically low. If the lateral diffusion of adatoms (adsorbed atoms) on the terrace is not sufficient, these adatoms build up to pyramid islands together with the appearance of steps. Such effects are further enhanced once islands are largely formed since adatoms cannot diffuse to and join existing islands. Consequently, the terrace-step-kink and pyramid structures predominate on the CVD Nb surface.

\begin{figure}[!t]
\centering
\includegraphics[width=\linewidth]{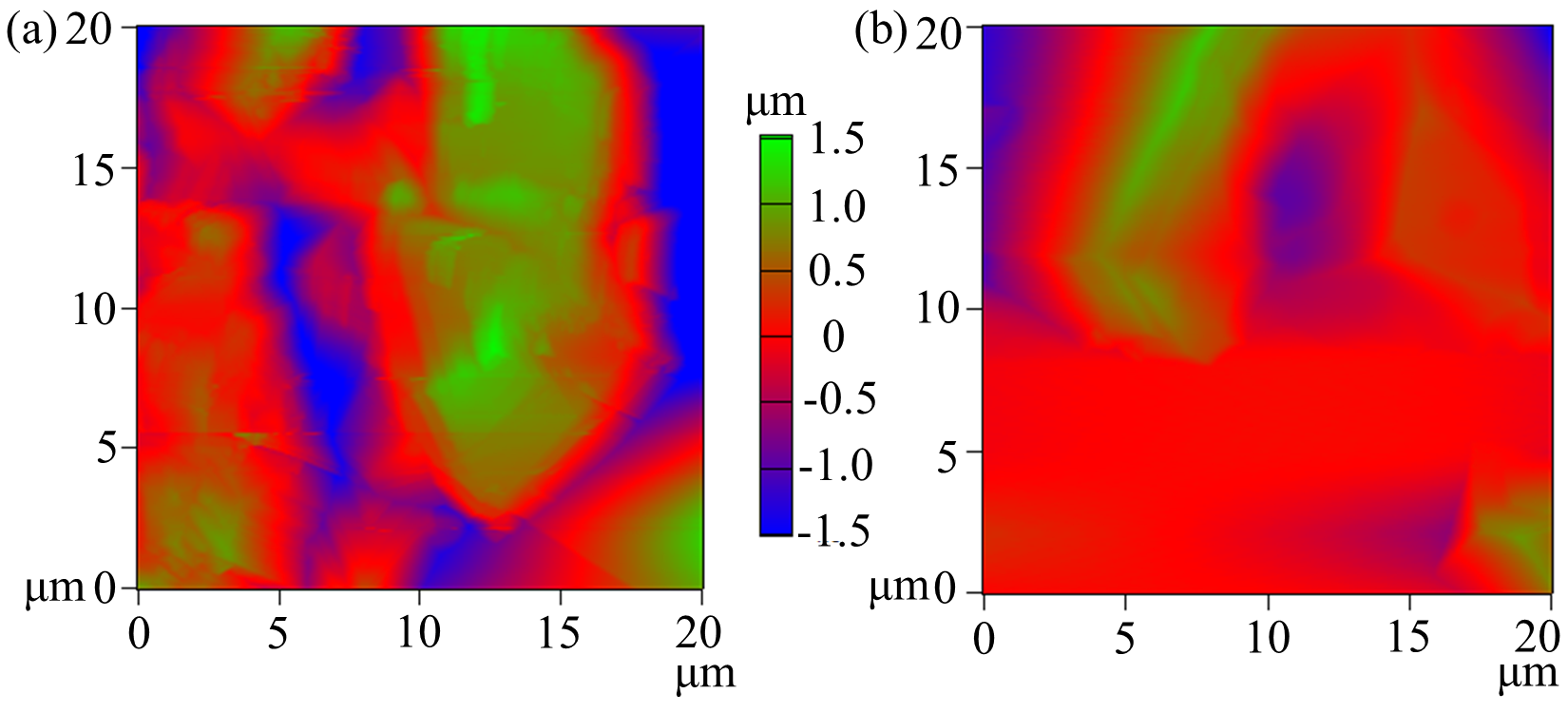}
\caption{Representative AFM images taken on the relatively flat regions (a) before and (b) after EP.}
\label{SunFig6}
\end{figure}

After CVD, EP polishing was conducted to alter the surface morphology regarding two aspects, $\textit{i.e.}$, removing or smoothing large pyramid structures, and eliminating surface steps and kinks. As demonstrated in Fig.~\ref{SunFig3}b and \ref{SunFig3}d, the edges and sharp features are greatly rounded after EP. Closer inspection of the cross-sections (Fig.~\ref{SunFig4}b) shows the regions that were relatively flat upon deposition are further smoothed; small islands are completely dissolved, while some large islands as high as 50 $\mu$m exist but their surfaces are also smoothed. This infers that kink and step sites, regardless of their locations, favor the onset of polishing, leading to a smooth and less-edged surface.

Due to the $\textit{ex situ}$ challenge, we compare the height of the highest pyramids observed before and after EP. For example, the pyramid height prior to polishing is as high as $\sim$ 100 $\mu$m, whereas the highest observed after polishing is $\sim$ 50 $\mu$m. This empirical comparison suggests the pyramids are polished by more than half in height, owing to intense macropolishing at these pyramids with a small radius of curvature (closer to R$_{2}$ in Fig.~\ref{SunFig2}a).

High-magnification images taken on the CVD pyramid (insert Fig.~\ref{SunFig4}a) show the pyramid consists of small nuclei (5 -- 10 $\mu$m) and exhibits a similar morphology of steps and kinks as other relatively flat regions. After EP (Fig.~\ref{SunFig4}b), these features disappear resulting in a smooth pyramid surface. This observation indicates micropolishing is also involved through leveling the height difference at steps and kinks and dissolving the small nuclei. Note that our primary motivation is to diminish the sharp features; while the existence of tall pyramids is not ideal, the smoothed pyramids would less severely impact the field enhancement. 

\begin{figure}[!t]
\centering
\includegraphics[width=8 cm]{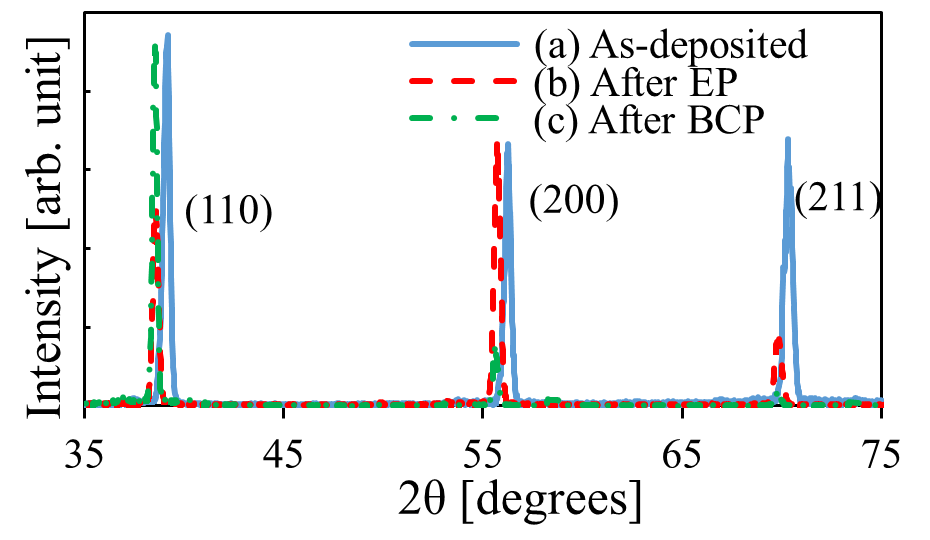}
\caption{XRD patterns of (a) as-deposited, (b) EP'ed, and (c) BCP'ed CVD Nb films. Intensities are normalized to their highest diffraction limit as referenced to as-deposited films.}
\label{SunFig7}
\end{figure}

\subsection{Surface roughness}

The quantification of surface roughness using AFM on a $>$10 $\mu$m uneven surface is challenging owing to the instrumental capability of the depth of field. The cross-sectional SEM images in Fig.~\ref{SunFig4} provide an empirical comparison of height change for pyramid structures before and after EP. Here, the AFM images were taken, as indications of roughness change, on the relatively flat regions.   

As shown in Fig.~\ref{SunFig6}, the smooth areas (denoted in red) are prominently enlarged after EP in the representative 20$^2$ $\mu$m$^2$ areas. Taking account of some inescapable small islands, the as-deposited samples have a large peak-to-valley distance of 4.2 $\mu$m. In contrast, the EP'ed samples exhibit a reduced value of 2.6 $\mu$m. Other surface parameters again indicate $\sim$ 50\% reduction of surface roughness, $\textit{e.g.}$, mean deviation (R$_{a}$) from 590 nm to 270 nm, and root mean square (R$_{q}$) from 740 nm to 390 nm. R$_{a}$ values from EP-smoothed regions on the film are close to the typical value ($\sim$ 300 nm) from an EP'ed bulk surface, which indicates the effectiveness of EP polishing when applied to thin films. Future work should focus on the removal of the remaining pyramid features.

\subsection{Crystal orientation}

\begin{figure}[!t]
\centering
\includegraphics[width=7.2 cm]{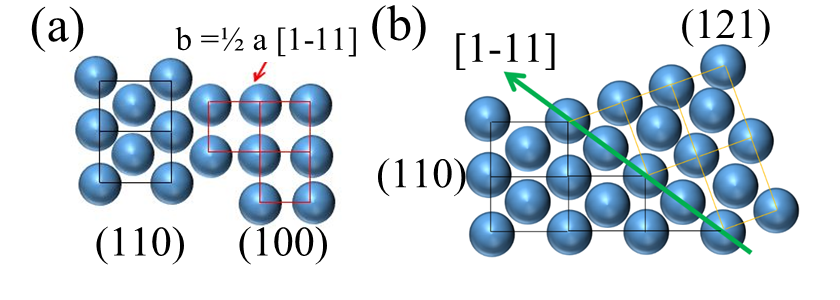}
\caption{Atomic models showing the formation mechanisms of (a) (100) and (b) (211) planes in addition to (110) planes. The lattice constant is denoted as “a”, and the Burgers vector is denoted as “b”.}
\label{SunFig8}
\end{figure}

The X-ray diffraction characteristics of electrochemically (EP) and chemically (BCP) polished CVD Nb films were compared (Fig.~\ref{SunFig7}). The as-deposited films exhibit a predominant (110) peak, epitaxy from the cubic Mo substrate, along with (100) and (211) diffractions. Fig.~\ref{SunFig8} illustrates the formation mechanisms of (100) and (211) planes in addition to the (110) epitaxy. In a body-centered cubic (bcc) structure, the [111] direction is the closest packed, and (110) planes could easily slip along this direction yielding (100) planes (Fig.~\ref{SunFig8}a). The Burgers vector of dislocations in between (100) and (110) planes is a/2 [111]. Additionally, rotating around the [111] axis by 70.5 degrees, the (211) and (110) planes can form the twin structure (Fig.~\ref{SunFig8}b). These twin structures are extensively observed under SEM which are marked by dashed lines in Fig.~\ref{SunFig3}a.

Moreover, we observed an orientation dependence during EP. For example, as shown in Fig.~\ref{SunFig7}, the highest diffraction peak changed to (100) planes from the initial highest (110) planes. Intensities were then normalized to that of (100) planes. Indeed, the (110) intensity reduced by half, and the (211) intensity likewise dropped exceeding half. (The shifting to smaller diffraction angles after EP indicates the compressive stress in the film is relieved.) 

The orientation-dependence behaviors, however, do pose some subtle questions for the conventional interpretation; the suppression of influences from crystal orientation is expected in micropolishing. In general, electropolishing is controlled by electrical, reaction, and diffusion processes. In micropolishing, the limiting factor nevertheless is the mass transport instead of charge transfer \cite{SunRef23}. The diffusion of species is a random motion and hence is believed to be orientation-independent, whereas the reaction-controlled polishing is typically orientation-dependent since the planer density that characterizes the average atoms in certain planes differs as summarized in Table \ref{table:1}.  

\begin{table}[h!]
\centering
\caption{Planer density and plane spacing of (110), (100), and (211) planes in Nb. The lattice constant (a) is 330 pm.}
\begin{tabular}{ |p {3.5cm} | p {1cm}| p {1cm} | p {1cm} |} 
  \hline
  Plane orientation & (110) & (100) & (211) \\ 
  \hline
  Planer density & $\frac{\sqrt{2}}{a^2}$ & $\frac{1}{a^2}$ & $\frac{\sqrt{6}}{3a^2}$ \\ 
  \hline
  Plane spacing & $\frac{\sqrt{2}a}{2}$ & $\frac{a}{2}$ & $\frac{\sqrt{6}a}{6}$ \\ 
  \hline
\end{tabular}
\label{table:1}
\end{table}

To test whether the orientation dependence during EP arises from a reaction-controlled process, we carried out BCP polishing that underwent similar chemical reactions as EP \cite{SunRef31}. From XRD (Fig.~\ref{SunFig7}), the (100) and (211) planes that have small planer densities show a pronounced reduction in intensity after BCP as compared to the (110) planes. This BCP behavior significantly differs from the EP results; it supports the theory that EP is less reaction-controlled. 

We further analyze the possible mechanisms that induce an orientation dependence. Our results have suggested that both macropolishing and micropolishing are involved in the EP process. Local electrical fields depending on geometry factors play a major role at the pyramids where local polishing-current densities are intensified resulting in large polishing rates. Upon assuming the statistical distribution of pyramids is uniform, the dominant population of (110)-structured pyramids are indicated by their highest intensity in as-deposited films (Fig.~\ref{SunFig7}a), and thus the global reduction of pyramids would exhibit a preference in the (110) plane. For example, comparing the pyramid cross-sections in Fig.~\ref{SunFig4}, the FWHM (full width at half maximum) remains the same value of 80 $\mu$m after EP, while the height reduces from 100 $\mu$m to 50 $\mu$m, suggesting the polishing substantially occurs in the perpendicular direction, say [110] orientation.  

Another possible mechanism is based on the conventional theory ($\textit{i.e.}$, mass transport controls EP); although the diffusion of species is orientation-independent, the oxide growth during EP (Fig.~\ref{SunFig2}b) varies in orientation. The large local polishing current produces thicker oxide layers and hence larger polishing rates -- this scenario would produce a similar outcome discussed above. Regardless of influences from the local polishing current, the oxide growth rate on the (110) plane is found to be higher than other planes \cite{SunRef33,SunRef34}. A thicker oxide layer on the (110) plane would induce a larger amount of removal on this plane during EP. Overall, preferential polishing is critical since it might provide selective polishing capabilities, and further investigations are necessary to confirm the mechanisms indicated by this work. 

\section{Conclusions}

In summary, electrochemical polishing (EP) was successfully performed on the chemical vapor deposited (CVD) Nb films to reduce the surface roughness, and compared with buffered chemical polishing (BCP). The characteristics of surface morphology, roughness, and crystal orientation have been analyzed to reveal the CVD growth and EP polishing mechanisms.

As-deposited films consist of relatively flat and pyramid-structured regions, which cause a large peak-to-valley distance of $>$ 100 $\mu$m. The observation of steps and kinks suggests that a terrace-step-kink model is responsible for the generation of pyramids. Also, the CVD crystals exhibit a large amount of (110) planes and some slip-induced (100) planes as well as the (211) twinning planes.

EP is demonstrated to effectively minimize the mean surface roughness on the relatively flat regions and significantly reduce the height of pyramids, $\textit{i.e.}$, by more than half. These smoothening behaviors are critical to enhancing the RF performance of CVD Nb-based cavities. Besides the reduction of pyramid height, the steps and kinks are found to disappear on the pyramids, indicating the involvement of both macroscale and microscale smoothing during the EP polish. The reaction-controlled mechanism is negligible in EP as suggested by a comparison with chemical polishing (BCP). The local enhanced current density and thickness variation of oxide dielectrics might be the controlling factors in the CVD-film polishing, leading to the crystal orientation dependence observed in this work. Overall, EP proceeds with more complex scenarios for CVD Nb films which contain the removal of both beyond and below-micrometer-scale sharp features. 

Our demonstration of the EP-CVD technology represents a viable application of Nb thin films for emerging superconducting applications.  

\section*{Data availability statement}

The data that support the findings of this study are available upon reasonable request from the authors.

\section*{Conflicts of interest}

V.A. and S.R.M. work at Ultramet. Z.S., M.G., J.T.M., and M.U.L. declare no competing financial interests.

\section*{Acknowledgments}
This work is funded by the U.S. Department of Energy SBIR phase-II award DE- SC0015727 and also supported by the National Science Foundation under Grant No. PHY-1549132, the Center for Bright Beams.

 \bibliographystyle{elsarticle-num} 
 \bibliography{maintext}





\end{document}